\documentstyle[aps,prl,floats,amssymb,graphicx]{revtex}

\begin{document}
\def\topfraction{1} \def\bottomfraction{1} \def\textfraction{0}

\twocolumn[\hsize\textwidth\columnwidth\hsize\csname @twocolumnfalse\endcsname

\title{Superconducting Nb-film LC resonator}
\author{A. Finne$^1$, L. Gr\"onberg$^2$, R. Blaauwgeers$^{1,3}$,
 V.B. Eltsov$^{1,4}$, G. Eska$^{5}$, \\ M. Krusius$^1$,
J.J. Ruohio$^1$, R. Schanen$^{1,6}$, I. Suni$^2$}

\address{$^1$Low Temperature Laboratory, Helsinki University of Technology,
P.O.Box 2200, FIN-02015 HUT, Finland \\
$^2$Microelectronics Centre, VTT Electronics, P.O.Box 1101,
FIN-02044 VTT, Finland \\
$^3$Kamerlingh Onnes Laboratory, Leiden University, P.O.Box 9504,
2300 RA Leiden, The Netherlands \\
$^4$Kapitza Institute for Physical Problems, Kosygina 2, 117334
Moscow, Russia \\
$^5$Physikalisches Institut, Universit\"at Bayreuth, D-95440
Bayreuth, Germany \\
$^6$CRTBT-CNRS, BP 166, F-38042 Grenoble Cedex 9, France}

\date{\today}
\maketitle

\begin{abstract}

Sputtered Nb thin-film LC resonators for low frequencies
at 0.5\,MHz have been fabricated and tested in the temperature range 0.05 --
1\,K in magnetic fields up to 30\,mT. Their $Q$ value increases towards
decreasing temperature as $Q \propto T^{-0.5}$ and reaches  $\sim 10^3$ at
0.05\,K. As a function of magnetic field $Q$ is unstable and displays
variations up to 50\,\%, which are reproducible from one field sweep to the
next. These instabilities are attributed to dielectric losses in the plasma
deposited SiO$_2$ insulation layer, since the thin-film coil alone reaches a
Q$\gtrsim 10^5$ at 0.05\,K.
\end{abstract}
 ]

\section{Introduction}

\begin{figure}[!!!t]
  \begin{center}
    \includegraphics[angle=0, width=8cm]{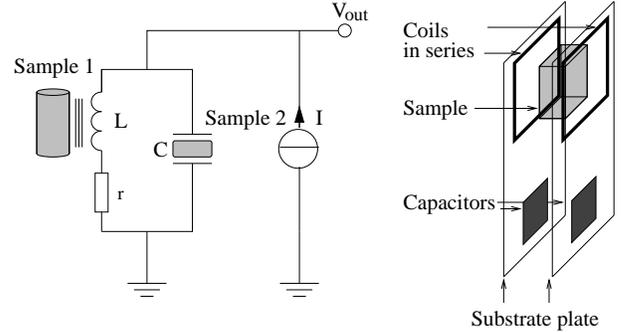}
\vspace{2mm}
    \caption{ {\it (Left)} Principle of the Q-meter measurement, where the
sample is coupled either inductively or capacitively to the resonator. {\it
(Right)}  The  sample is sandwiched between two thin-film devices. In our
magnetic measurements the sample is placed between two planar coils, which are
connected in series, while the capacitors are fabricated as complete thin-film
components on the 0.5\,mm thick wafer, as shown in
Fig.~\protect\ref{fig:design}.}

    \label{fig:idealsetup}
  \end{center}
\end{figure}

An electrical resonator with a high quality factor $Q$ is a sensitive device
for the measurement of materials' properties. At low frequencies (up to several
MHz) the inductive and capacitive elements are spatially separated and allow
well-defined characterization of the magnetic or dielectric properties of the
sample piece. Such resonators can be assembled from discrete components,
i.e.~a coil wound from superconducting wire and a high-$Q$ capacitor
\cite{Ruutu}. With a parallel-coupled LC resonator,
$Q$-values up to $9\times10^4$ have been reached at temperatures below 0.1\,K,
when the solenoidal coil is wound from $25\,\mu$m Nb wire, and the resonator is
directly coupled to the gate of a GaAs FET amplifier which is operated in a
4\,K environment \cite{Koivuniemi}. In our application the resonators have
been employed for low-frequency continuous-wave NMR
\cite{3HeNMR}. In most measuring applications a more efficient geometry for
the resonators is a planar Nb thin-film construction where the inductively or
capacitively coupled sample is sandwiched between two thin-film devices
(Fig.~\ref{fig:idealsetup}).

The sensitivity of the measurement is often determined by the intrinsic losses
of the resonator, i.e.\ by its unloaded $Q_0$ value. The magnetic
susceptibility $\chi_m$ of the sample changes the inductance
\begin{equation}
  \label{eq:l}
  L=L_0(1+\chi_m)\quad,
\end{equation} or equivalently in a capacitive measurement, the dielectric
susceptibility $\chi_e$ is seen in the capacitance
\begin{equation}
  \label{eq:c}
  C=C_0\epsilon_0(1+\chi_e)=C_0\epsilon\quad.
\end{equation} The frequency-dependent complex susceptibilities,
$\chi=\chi'-i\chi''$, consist of the dispersion
$\chi'(\omega)$ and absorption $\chi''(\omega)$. In the case of dielectrics,
the dielectric constant is usually used:
$\epsilon=\epsilon'-i\epsilon''$. $L_0$ and $C_0$ are constants which are
defined by the thin-film component design. Usually the sample does not
fill the entire active volume, which is accounted for by
introducing a filling factor, to modify the effective susceptibility (see
e.g.\ \cite{abragam}).

In Fig.~\ref{fig:idealsetup} the resonator is driven from a constant current
source and the output voltage is proportional to the impedance
$Z(\omega)$ of the parallel LC circuit. The width of the resonance peak is
determined by the losses, while  the resonance frequency $\omega_r$ depends on
the real component of the susceptibility $\chi'$. Both quantities can be
extracted by recording the
transfer function $Z(\omega)$ as a function of $\omega$ and by fitting the
measured curve to the expected impedance of the equivalent circuit.

Frequently applications are in the limit where $\chi(\omega)$ is small and one
may use linear response at resonance. Then the sensitivity of the measurement
improves with increasing $Q$. This limit is usually applied in continuous wave
NMR measurement, where the resonator is driven at resonance and its output is
modulated by taking the sample through magnetic resonance, by sweeping an
externally applied magnetic field $H$. Dispersion and absorption are then
extracted from the resonance response
\cite{abragam}
\begin{equation}
  \label{eq:zind}
  Z=\frac{\omega_0L_0Q_0}{1+iQ_0\chi(H)}\quad.
\end{equation} In the case of a dielectric sample the frequency shift due to
$\epsilon'$ is typically large and requires retuning of the resonator from
$\omega_0$ to $\omega_r = (LC_0\epsilon')^{-1/2}$. If we rewrite the
dielectric constant in the form $\epsilon=\epsilon'+
\delta\epsilon'-i\epsilon''$, where $\delta\epsilon'$ represents small changes
in $\epsilon'$ during the measurement, then the expression analogous to
Eq.~(\ref{eq:zind}) becomes
\begin{equation}
  \label{eq:zel} Z=\frac{\omega_rL_0Q_0}{1-iQ_0(\frac{\delta\epsilon'}
{\epsilon'}-i\tan\delta)}\quad,
\end{equation} where $\tan\delta \approx \epsilon''/\epsilon'$ is the loss
tangent and $Q_0$ is assumed large.

\section{Design and fabrication}

\begin{figure}[!t]
  \begin{center}
    \includegraphics[angle=0, width=8cm]{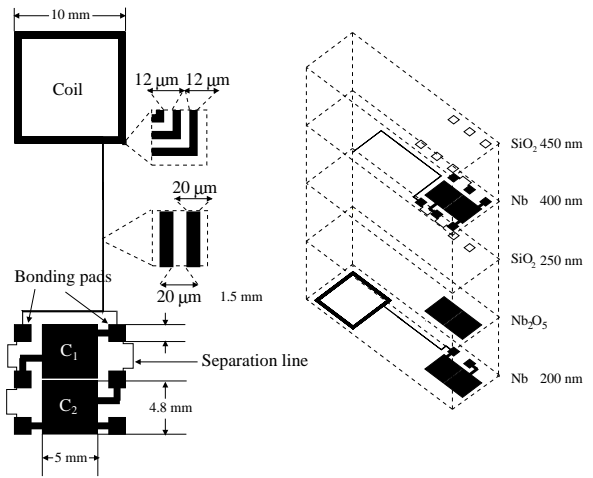}
\vspace{2mm}
    \caption{Layout and dimensions of the resonator {\it (left)}, and the
elements in different thin-film layers {\it (right)}.  The Nb$_2$O$_5$ layer
is applied only to some of the resonators.}
    \label{fig:design}
  \end{center}
\end{figure}

Our thin-film resonator design is shown in Fig.~\ref{fig:design}. These
devices are meant for magnetic measurements where the sample is placed between
two series-coupled planar coils on separate substrates, in a
Helmholtz-like configuration.  A static polarizing magnetic field is applied
in the direction parallel to the two resonator plates.

The resonators are fabricated on $\varnothing\, 100\,$mm silicon and sapphire
wafers. The silicon wafers are cut from usual micro-fabrication grade n-doped
single crystal material, with a 240\,nm thick thermally oxidized SiO$_2$
surface. The [11\=20] sapphire wafers, with an epitaxially polished surface,
are used as reference substrates, to compare to a pure material with
inherently low losses. The resonator circuit is fabricated in four layers
which are summarized in Table~\ref{tab:layer}. In addition, a layer of
Nb$_2$O$_5$, made by anodizing Nb in the lower Nb layer, is added to
some capacitors, to test the properties of this dielectric.

The Nb layers are sputter deposited. The superconducting transition
temperature of three 200\,nm thick Nb test layers was measured to be 9.12 --
9.20\,K, with a width of the transition region $\lesssim 5\,$mK. The SiO$_2$
insulation and passivation layers are deposited with plasma enhanced
chemical vapor deposition (PECVD) from silane SiH$_4$ gas. The first 50\,nm
are deposited at 120\,$^\circ$C and the rest at 150\,$^\circ$C. This process
leaves typically a layer, which contains impurities such as OH$^-$ radicals,
and its etch rate in HF acid is high.

\begin{table}[b]
  \begin{center}
    \begin{tabular}{c|l c c l}  Layer & Function   & Material    & Thickness &
Fabrication \\
      & &           & nm         & method      \\ \hline

4     & Protection & SiO$_2$     & 450       & PECVD       \\  3     &
Conductor & Nb          & 400       & Sputtering     \\  2     & Insulation &
SiO$_2$     & 250       & PECVD       \\ \hline\hline  1b    & Insulation &
Nb$_2$O$_5$ & 140       & Anodization \\  1b    & Conductor  & Nb          &
150       & Sputtering     \\ \hline

1a    & Conductor  & Nb          & 200       & Sputtering

    \end{tabular}
\vspace{2mm}
    \caption{Consecutive thin-film layers (applied in the order
$1\rightarrow4$), their thickness, and application method. Layers (1a) and
(1b) are alternative and refer to different types of resonators. In (1b) a
200\,nm layer of Nb is sputtered, as in (1a), but it is then anodization
oxidized to a depth of about 50\,nm, to form a 140\,nm thick Nb$_2$O$_5$
layer. It is the SiO$_2$ insulation layer (2) which is identified as the
source of the dielectric losses in the Q-value measurements. }
    \label{tab:layer}
  \end{center}
\end{table}

The dimensions of the resonator are shown in Fig.~\ref{fig:design}. The coil
consists of 25 turns, with a linewidth and line spacing of 12\,$\mu$m. The
coil was measured to have an inductance of 17\,$\mu$H at 4\,K. The capacitor
consists of a SiO$_2$ layer ($\epsilon_r\sim 3.8$ at 4\,K), sandwiched between
two Nb films. The total capacitance of 8\,nF is composed of two identical
capacitors, connected with lines which can later be cut, similar to the
connection between the capacitance and inductance. This allows to sample the
uniformity of the fabricated components and to operate the resonator in
different configurations. The bonding pads are designed to be large, to allow
easy electrical connections.

\begin{figure}[!t]
  \begin{center}
    \includegraphics[angle=0, width=7cm]{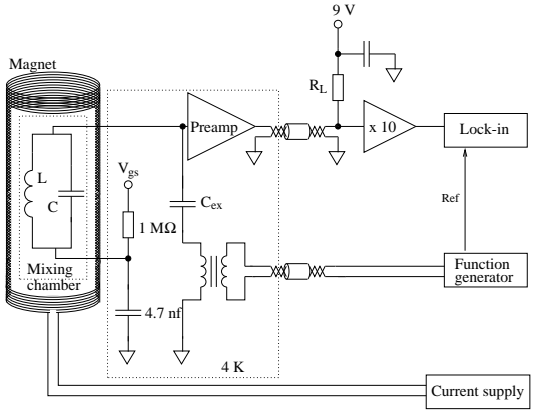}
\vspace{2mm}
    \caption{Measuring circuit. The resonator is excited with a function
generator and the response is recorded with an amplifier chain consisting of a
wide-band FET preamplifier in the liquid He bath, plus of a narrow
band amplifier and a lock-in amplifier at room temperature. }
    \label{fig:measetup}
  \end{center}
\end{figure}

Before use, the resonators are screened in room-temperature resistance
measurements for fabrication failures. Less than 10\,\% of the capacitors
show very low resistance and are assumed short circuited. The expected
value for the coil resistance is 106\,k$\Omega$ at room temperature. However,
only about 20\,\% of the coils reach this value. Others have a reduced
resistance indicating that a parallel shunt resistance exists across some
turns in the coil. In this respect the most delicate section of the present
design is the
return lead from the center of the coil in the top most Nb layer. This lead
crosses all 25 turns of the coil winding, with 50 steps in layer height. In
most cases coils with a reduced resistance at room-temperature recover the
proper value of 30\,k$\Omega$ on cooling to 77\,K. To cure
problems with the SiO$_2$ insulation, some resonators were annealed at
$1000^{\circ}$C for 2\,h. This turned out detrimental to the Nb film. Other
annealing efforts at a lower temperature of $800^{\circ}$C were equally
unsuccessful. Otherwise the resonators are durable, both with respect to
storage at room temperature and thermal cycling to 4\,K.

\section{Measurement setup}

The measuring setup is shown in Fig.~\ref{fig:measetup}. The resonator is
placed inside the mixing chamber of a small $^3$He--$^4$He dilution
refrigerator which cools to 50\,mK. To minimize losses, the resonator
is housed in an extension of the mixing chamber which is machined from
araldite epoxy. A heat treated high-conductivity copper shield is inserted
between the mixing chamber and the steel vacuum jacket, to reduce dissipation
in the metal parts. A superconducting solenoid outside the vacuum jacket in
the liquid He bath is used to generate a homogenous magnetic field parallel to
the Nb film.

Ultrasonic bonding with $\varnothing\, 25\,\mu$m Al alloy wire is used to
connect to the bonding pads on the resonators. High $Q$ values are achieved,
even when the bonding wire is part of the resonance circuit with two Nb-film
coils connected in series. This indicates good contact of the bonding wire to
the Nb pad. Unfortunately the  Al alloy loses its superconductivity at a low
magnetic field ($\approx 8$\,mT).
\begin{figure}[!t]
  \begin{center}
    \includegraphics[angle=0, width=8.5cm]{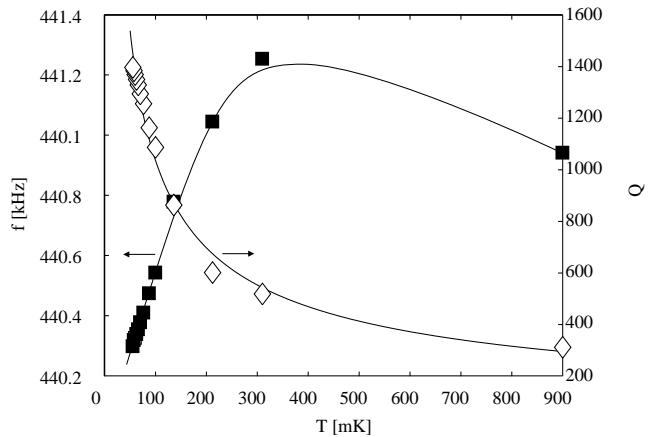}
\vspace{2mm}
    \caption{Temperature dependences of $Q$ ($\diamond$) and
$f=\omega_r/(2\pi)$ ($\blacksquare$) of a thin-film resonator on
silicon substrate (no anodization in capacitors). The fitted curve
represents the temperature dependence $Q\propto T^{-0.5}$. The same
results were measured for resonators on sapphire substrate. }
    \label{fig:tfq}
  \end{center}
\end{figure}

The voltage across the resonator is measured with a FET preamplifier operating
at 4\,K in the liquid He bath, similar to that described in Ref.~\cite{Ruutu}.
The resonator is connected to the preamplifier via a twisted pair of
superconducting wires. The resistance in these leads outside the resonator
loop is not critical since most of the current flow occurs within the
resonator. The resonator is  excited via a transformer and a coupling
capacitance $C_{{\rm ex}}$. This represents  a nonideal current source,
especially for a high-$Q$ resonator, but it can be shown to be equivalent to
an ideal source if the resonator capacitance $C$ is replaced with an effective
capacitance $C+C_{{\rm ex}}$. Thus the feeding capacitor $C_{{\rm ex}}$ needs
to be small with high $Q$, not to load the resonator.

The liquid-He-temperature preamplifier is an integral part of the resonator
since measurements with $Q$ values approaching $10^5$ are not possible without
a high-input-impedance device. Even so, we find that the input FET has a leak
resistance and parasitic capacitance which load the resonator at the highest
impedance levels. In the present measurements the total noise reduced to the
input of the preamplifier is $\sim 4\,\rm{nV}/\sqrt{\rm{Hz}}$.

\section{Results on $Q$ values}

In Fig.~\ref{fig:tfq} the temperature dependences of $Q$ and the resonance
frequency $\omega_r$ are shown for a single resonator of the type shown in
Fig.~\ref{fig:design}, with only SiO$_2$ insulation in the capacitors. The
$Q$ values turned out to be low, (1 -- 2)$\times 10^3$ at 50\,mK, on both
silicon and sapphire substrates, and within measuring precision did not
depend on the resonance excitation level in the regime 0.01 -- 1\,mV across
the resonator (at constant applied magnetic field). The resonance
frequency has a maximum at about 400\,mK, which corresponds to a minimum
in the dielectric constant of the SiO$_2$ insulation in the capacitors.

To resolve the origin of the low $Q$ value, the thin-film components were
tested separately. If the thin-film coil on the resonator plate is replaced
with an external wire-wound Nb coil with a similar inductance and high $Q$
value, then the resonator $Q$ is not significantly changed from those recorded
in Fig.~\ref{fig:tfq}. In contrast,
\begin{figure}[!t]
  \begin{center}
    \includegraphics[angle=0, width=8cm]{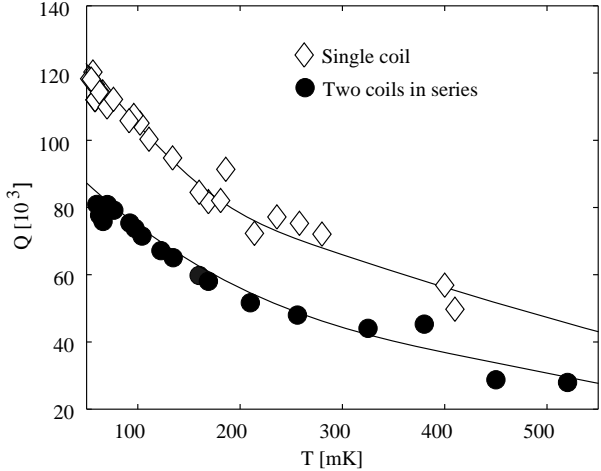}
\vspace{2mm}
    \caption{$Q$ values of thin-film coils on silicon substrate connected in
parallel with external
high-$Q$ capacitors measured at 440\,kHz. The upper curve
$(\Diamond)$ applies for a single thin-film coil while the lower $(\bullet)$
describes a series connection of two coils.}
    \label{fig:highq}
  \end{center}
\end{figure}
when the thin-film capacitance on the resonator plate is replaced with
external low-loss capacitors \cite{High-Q-Capacitor}, much higher $Q$
values are
measured, as shown in Fig.~\ref{fig:highq}: $Q$ climbs at 50\,mK to
$1.2\times 10^5$ with a single thin-film coil and to $0.8\times 10^5$ with two
coils on different resonator plates coupled in series with Al bonding
wires. Thus the thin-film coils perform well, at least in terms of their $Q$
value. On warming to 500\,mK, the $Q$ value drops by a factor of 3, which is
approximately the same behavior as for the complete thin-film resonators in
Fig.~\ref{fig:tfq}. With external capacitances the change in resonance
frequency is small, $\Delta\omega_r /(2\pi) \sim 1\,$Hz, compared to $\sim
1\,$kHz in Fig.~\ref{fig:tfq}. From these test we conclude that the losses
of the resonator are dominated by the SiO$_2$ insulation layer in the
capacitors.

\begin{figure}[!t]
  \begin{center}
    \includegraphics[angle=0,width=8.5cm]{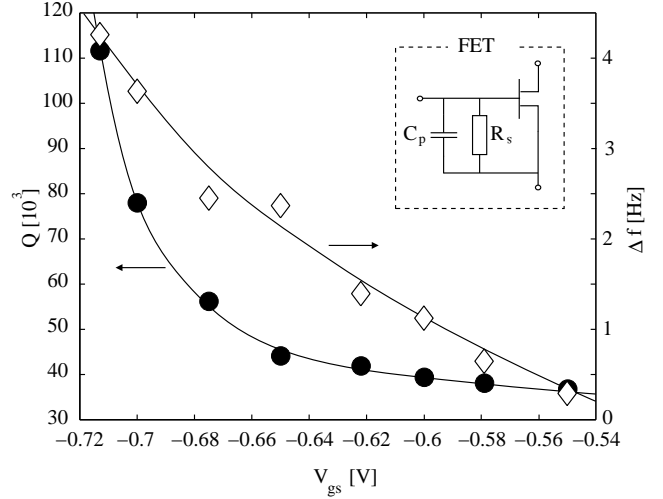}
\vspace{2mm}
    \caption{Dependence of the measured $Q$ value ($\bullet$) and
the change in resonance frequency $\Delta f$ ($\Diamond$) on the
gate-to-source bias voltage $V_{gs}$ (Fig.~\protect\ref{fig:measetup}) of the
input FET of the preamplifier in the liquid He bath. When
 $V_{gs}$ is reduced below -0.72\,V, the circuit becomes unstable and starts
oscillating. A single thin-film coil on silicon substrate is used here at
50\,mK and 445\,kHz, coupled in parallel with external capacitors.}

    \label{fig:vgs}
  \end{center}
\end{figure}

Finally, it is instructive to note that the high $Q$ values in
Fig.~\ref{fig:highq} are sensitive to the preamlifier bias settings.
Fig.~\ref{fig:vgs} shows the dependence on the gate-to-source voltage of the
input FET. An equivalent circuit for the FET input \cite{FET} is shown in
the inset where the input is replaced with a parasitic capacitance C$_p$ and
a shunt resistance $R_s$ which both are connected in  parallel to the
resonator. Their values depend on the FET bias settings.
\suppressfloats
The change in capacitance, as calculated from $\Delta \omega$ in
Fig.~\ref{fig:vgs}, corresponds to 0.1\,pF. To obtain the measured
$Q=1.2\times 10^{5}$, the input resistance would be
$R_s \approx 11\,$M$\Omega$ in the case of an ideal resonator. Since the
resonator is not
ideal, the FET input resistance is higher. Assuming, as an example, the
resonator $Q_0$ to be $2\times 10^5$, then the input resistance of the FET
needs to be 30\,M$\Omega$, to produce a loaded $Q$ of 1.2$\times 10^{5}$.

\section{Magnetic field effects}

\begin{figure}[!t]
  \begin{center}
    \includegraphics[angle=0, width=8.5cm]{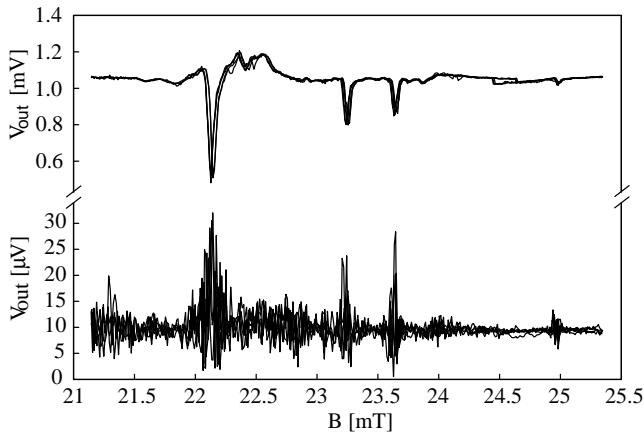}
\vspace{2mm}
    \caption{Examples of unstable resonator operation as a function of the
externally applied magnetic field intensity $B$ at 90\,mK and
440\,kHz. The output voltage $V_{out}$ of the preamplifier is
monitored at two different amplitudes of resonance excitation: The
peak voltage in the thin-film capacitors across the SiO$_2$ insulation
corresponds to 1\,mV {\it (top)} and 10\,$\mu$V/m {\it (bottom)}.  The
resonator is fabricated on sapphire substrate. In both panels the
magnetic field is swept twice back and forth. Apart from the change in
magnetic field dependence, the ac electric field does not change the
average $Q$ value significantly at excitation levels in the 0.01 --
1\,mV range.}
    \label{fig:instab}
  \end{center}
\end{figure}

The resonators turned out to be highly sensitive to applied magnetic
field. Fig.~\ref{fig:instab} shows an example where the voltage across the
resonator is recorded at resonance as a function of magnetic field. The
resonance frequency is not substantially changing during the measurement, thus
the changes in output voltage are due to absorption and correspond roughly to
similar changes in $Q$ value. The output is not reproducible from one
resonator to the next, but reproduces from one measurement to the next as a
function of field for one particular resonator. At low excitation (bottom
panel) the  changes in output voltage as a function of field  have the
appearance of noise, with both positive and negative changes from the average.
At higher excitations (top panel) the changes look more like reductions in $Q$
value. The output appears to be independent of the sweep rate of the magnetic
field and if the sweep is stopped, then the last value is preserved. The
unstable behavior continues all the way to zero magnetic field.

\suppressfloats

The magnetic field dependence of $Q$ is present in similar magnitude in
resonators on both silicon and sapphire substrates, also with and without the
extra Nb$_2$O$_5$ insulation. If the thin-film coil is replaced with a
high-$Q$ wire-wound coil, then the output as a function of field remains
unchanged. In contrast, a thin-film coil connected in parallel with external
high-$Q$ capacitors produces a much more stable output as a function of field
(at a much increased $Q$ value). However, the thin-film coil also suffers from
the magnetic field dependence, although to a smaller extent. For instance,
with the two thin-film coils in series in Fig.~\ref{fig:highq}, the changes in
output voltage are $\sim2\,$\% at 62\,mK, where $Q = 8\times10^4$. In
addition, the thin-film coil exhibits a monotonous drop in Q value as a
function of  magnetic field. A measurement on the series connection of two
thin-film coils at 63\,mK showed a
$Q=6\times10^4$ at zero field which dropped by about 10\% when the field was
increased to 7\,mT, while the resonance frequency remained unchanged to within
1\,Hz. This is a sufficiently small change that it would not be measurable for
the complete resonator, where the low $Q$ value is determined by the
capacitors. In the last example the bonding to the coils was done with
Al alloy wire. This results in a drastic drop in Q when the bonding wire
loses its superconductivity above 8\,mT.

Our measurements suggest that the unstable magnetic field dependence
originates primarily from the SiO$_2$ insulation: The effects are strong when
the thin-film capacitors are used. The thin-film coil has a small parasitic
capacitance of about 0.3\,pF, which also involves the SiO$_2$ insulation:
This capacitance arises when the lead from the center of the coil crosses
over all the turns of the planar windings. Thus the
magnetic field effects of the complete thin-film resonator and those of the
thin-film coil alone are scaled roughly in proportion to their capacitances
with SiO$_2$ insulation. Similar magnetic-field-dependent effects have been
observed previously in thin-film coils of much smaller size \cite{minicoils}.

\section{Amorphous thin-film insulator}

Both the reduced Q of the resonators and their unstable losses as a
function of magnetic field have here been traced to originate
primarily from the amorphous PECVD-deposited SiO$_2$ insulation in the
capacitors. Amorphous dielectrics are known to be lossy and to exhibit
as a function of temperature a minimum in the dielectric constant at a
few hundred mK \cite{pobell}, as was the case here. Recently the
dielectric constant of the multicomponent glass
BaO-Al$_2$O$_3$-SiO$_2$ has been reported to be sensitive to magnetic
fields at temperatures of a few mK \cite{magfield,magglass}. These
measurements were conducted at 1\,kHz, whereas here we operate at a
higher frequency of 400\,kHz. The dielectric constant has also been
found to be sensitive to the excitation amplitude \cite{osheroff}.
With our low excitation levels, we did not find significant dependence
of the average dielectric losses or the resonance frequency on the
applied ac electric field in the measurements of Figs.~\ref{fig:tfq}
or \ref{fig:highq}.

In addition to the amorphous structure of the insulation layer,
impurities, or the interface between the dielectric and the Nb may
also contribute to the observed effects.

\section{Conclusions}

The present work shows that Nb-film fabrication methods turn out a good yield
of durable resonators, and that the
sputtered Nb film is a high-quality conductor on a standard
doped silicon substrate at frequencies of about 1\,MHz at mK temperatures. In
contrast, PECVD deposited SiO$_2$ insulation does not reach corresponding
quality levels: Its dielectric losses are large and highly
magnetic field dependent, even around zero field. Additional anodized
Nb$_2$O$_5$ insulation did not increase the losses and might be of
better quality.

Secondly, the thin-film inductor is of sufficiently high quality such that it
can be used to investigate the properties of the dielectric in the capacitor,
to find better alternatives. The resonator method with a high-Q coil and a GaAs
MESFET preamplifier thereby allows access to the frequency range from 0.1\,MHz
up to several MHz, which so far has been only rarely investigated in studies of
amorphous solids.

\section*{Acknowledgements} This work was funded in part by the EU-IHP
programme (contract no.~HPRI-1999-00098). We thank the personnel at the
Microelectronics Centre for their help with the fabrication of the
resonators.

\end{document}